\documentclass[preprint,12pt,authoryear]{elsarticle}

\usepackage{amssymb}
\usepackage{caption}
\usepackage{tabularx} 
\usepackage{array}
\usepackage{multirow}
\usepackage{tikz}
\usepackage{booktabs}
\usepackage{hyperref}
\hypersetup{
    pdfborder={0 0 0} 
}

\newcommand{\circled}[1]{\tikz[baseline=(char.base)]{
   \node[shape=circle,draw,inner sep=0.1pt] (char) {#1};}}

\makeatletter
\def\ps@pprintTitle{%
 \let\@oddhead\@empty
 \let\@evenhead\@empty
 \def\@oddfoot{}%
 \let\@evenfoot\@oddfoot
}
\makeatother

\begin{document}

\begin{frontmatter}



\title{Novobo: Supporting Teachers' Peer Learning of Instructional Gestures by Teaching a Mentee AI-Agent Together}


\author{Jiaqi Jiang\textsuperscript{a}\corref{cor1}} 
\author{Kexin Huang\textsuperscript{a}\corref{cor1}}
\author{Roberto Martinez-Maldonado\textsuperscript{b}}
\author{Huan Zeng\textsuperscript{c}}
\author{Duo Gong\textsuperscript{a}}
\author{Pengcheng An\textsuperscript{a}\corref{cor2}}
\ead{anpengcheng88@gmail.com}

\cortext[cor1]{Co-first Authors}
\cortext[cor2]{Corresponding Author}

\affiliation{organization={Southern University of Science and Technology},
            addressline={Xueyuan Blvd},
            city={Shenzhen},
            postcode={518055},
            state={Guangdong},
            country={China}}

\affiliation{organization={Monash University},
            addressline={20 Exhibition Walk, Clayton},
            city={Melbourne},
            postcode={3800},
            state={Victoria},
            country={Australia}}

\affiliation{organization={Beijing Normal University},
            addressline={18th Jinfeng Road, Zhuhai},
            city={Zhuhai},
            postcode={519087},
            state={Guangdong},
            country={China}}

\fntext[]{See a video demonstration via: \href{https://drive.google.com/file/d/1Jaa97N1ZClGF_61Kkbr4taTbJ2hRxrW9/view?usp=sharing}{\textcolor{blue}{Click Here}}}

\begin{abstract}

Instructional gestures are essential for teaching, as they enhance communication and support student comprehension. However, existing training methods for developing these embodied skills can be time-consuming, isolating, or overly prescriptive. 
Research suggests that developing these tacit, experiential skills requires teachers’ peer learning, where they learn from each other and build shared knowledge. 
This paper introduces Novobo, an apprentice AI-agent stimulating teachers' peer learning of instructional gestures through verbal and bodily inputs.
Positioning the AI as a mentee employs the learning-by-teaching paradigm, aiming to promote deliberate reflection and active learning.
Novobo encourages teachers to evaluate its generated gestures and invite them to provide demonstrations. 
An evaluation with 30 teachers in 10 collaborative sessions showed Novobo prompted teachers to share tacit knowledge through conversation and movement. 
This process helped teachers externalize, exchange, and internalize their embodied knowledge, promoting collaborative learning and building a shared understanding of instructional gestures within the local teaching community.
This work advances understanding of how teachable AI agents can enhance collaborative learning in teacher professional development, offering valuable design insights for leveraging AI to promote the sharing and construction of embodied and practical knowledge.
\end{abstract}


\end{frontmatter}


\section{Introduction}
\label{sec:intro}


In classroom teaching, nonverbal communication plays a crucial role in conveying emotions, intentions, and content between teachers and learners through cues such as gestures, facial expressions, and postures \citep{abakumova2021non}. 
Among all these nonverbal behaviors, instructional gestures have been widely recognized and validated by educational studies for their profound impact in classroom teaching \citep{ goldin2011learning, roth2001gestures, smotrova2017making, novack2015learning}.
For example, teachers' gestures contribute to enhancing their expressiveness \citep{kita2000representational}, affect learners' emotions, and can facilitate learners' comprehension of specific topics \citep{abakumova2021non, kelly2010two}. 
However, despite the significant role of instructional gestures as highlighted in the educational literature above, the development of effective instructional gesturing skills often receives relatively little attention in Teacher Professional Development (TPD) \citep{bambaeeroo2017impact, van2012makes, klinzing1987training}.

A common approach to training nonverbal behaviors of teachers involves video recording instructional segments from experienced teachers and showing them to novice teachers to review teaching practices \citep{calandra2009using, rich2009video}. 
However, this method relies on extensive video analysis and editing work, which requires expert support, thereby incurring significant time and effort costs \citep{hamidah2019video}. 
Another method for training teachers' nonverbal behaviors is based on behavior modification theory \citep{mcdonald1973behavior}. 
For instance, \citet{barmaki2015providing} utilized an approach based on machine learning to automatically classify teacher instructional gestures into two categories---open or closed---and correspondingly showing positive or negative feedback.
While such methods can effectively train specific aspects of nonverbal teaching behaviors, the evaluation criteria they employ could be rather simplified and prescriptive, overlooking the diversity and individuality of teachers' nonverbal behaviors \citep{klinzing1987training}. Moreover, AI training systems which limit feedback to binary outcomes, such as correct or incorrect, without providing any explanation for the reasoning behind these decisions, can hinder teachers’ understanding and effective application of the feedback \citep{gajos2022people}.

Gesture based nonverbal behaviors represents a form of tacit knowledge that is acquired in practice and difficult to convey through textual or verbal means \citep{westerlund2021s}. 
Learning instructional gestures often occurs through collaborative learning with colleagues or expert teachers, with peer feedback being an important context for this process \citep{seroussi2019reflections}.
In such scenarios, teachers become active participants in knowledge construction, exploring issues, sharing insights, and building new understanding and knowledge through mutual feedback \citep{hanrahan2001assessing}. 
In offering peer feedback and peer assessments, both the provider and recipient benefit: the provider enhances their analytical and reflective skills, while the recipient gains new insights and areas for improvement \citep{Hattie2007}. 
Consequently, research advocates for the training of teachers' nonverbal behaviors through collaborative learning methods among teachers, where they can teach and learn from each other \citep{khuman2024impact}. 
However, the avoidance of criticizing others, which could lead to unpleasant situations \citep{seroussi2019reflections}, could result in biases in the reliability of peer assessments \citep{brown2004teachers}, such as ``friendship marking'' or may lead to feedback being provided in the form of neutral or vacuous ``social'' comments \citep{xie2008effect}.
This can diminish the effectiveness that peer assessment could potentially have.
To fully leverage the advantages of peer learning while mitigating the negative aspects of avoiding direct criticism, a promising approach is the introduction of a neutral, virtual agent acting as a mentee. This allows teachers to focus on instructing and explicitly externalizing their instructional behaviors without the hesitation or concern caused by social factors --- through the use of Teachable Agents (TAs) \citep{johansson2023intelligent}. 
The development of Large Language Models (LLMs) offers an opportunity to create low-cost, natural, and intelligent agents \citep{bhowmik2024evaluation}.
TAs have the potential to help maximize the benefits of peer feedback by creating a structured environment where teachers can engage in shared reflection and interaction, focusing on the virtual agent, thus minimizing feelings of criticism and enhancing productive collaboration.

TAs have been typically designed based on the \textit{Learning by Teaching} (LBT) approach, where learners instruct the TAs, thereby articulating, restructuring, and reflecting on their knowledge \citep{roscoe2007understanding, song2017designing}. 
Furthermore, it has been explored that when learners collaboratively teach an agent, it promotes discussion among them, ultimately leading to the integration of different viewpoints \citep{yu2009teachable}.
Although TAs have typically been designed for students, their potential to promote peer knowledge exchange offers a valuable opportunity for fostering peer learning and feedback practices among teachers.

\begin{figure*}[h]
    \centering
    \includegraphics[width=\textwidth]{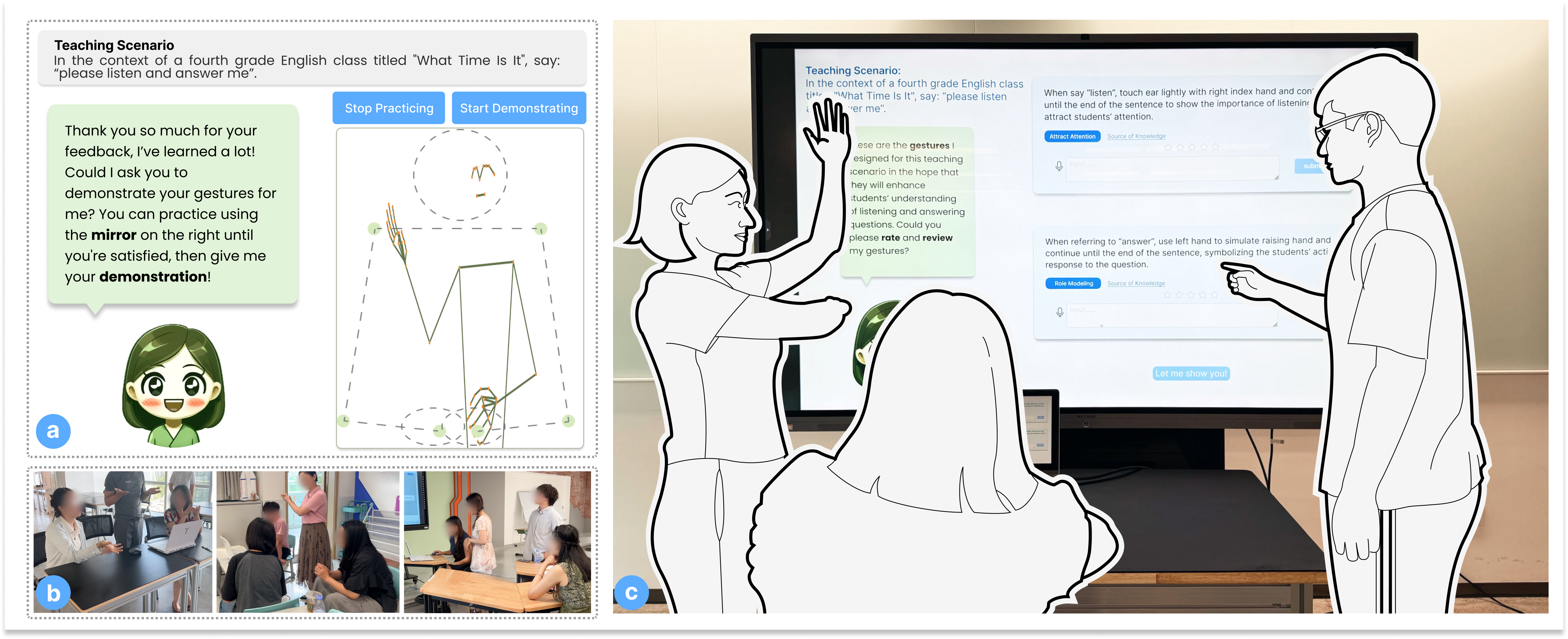}
    \captionof{figure}{(a) Demonstration stage of the Novobo interface; (b) Teachers using Novobo during field evaluation; (c) Typical session scenario of a teacher study group.}
    \label{fig:teaser}
\end{figure*}

We present Novobo, a teachable AI-powered agent that acts as an apprentice, novice teacher, seeking teachers' collective mentoring on instructional gestures. 
It is targeted at the context of teachers' study groups, a key setting in professional development where they share knowledge and learn together.
Novobo is designed to stimulate peer learning through both verbal and bodily inputs from teachers (see a typical interaction scenario in Figure \ref{fig:teaser}). 
The interaction with Novobo repeats a four-step loop: (1) \textit{Posing Questions}: Teachers discuss and assign a teaching scenario, asking Novobo to propose suitable instructional gestures and reviewing its answer; (2) \textit{Commentary}: Teachers provide collective ratings to Novobo's proposed ideas, along with detailed commentaries; (3) \textit{Demonstration}: Novobo asks teachers to demonstrate a suitable gesture in front of a mirrored body outline which they can practice with and record movement once satisfied. After recording, teachers can replay to review their instructional gestures and make any necessary adjustments; (4) \textit{Explanation}: Teachers give a verbal description of the demonstrated gesture and collectively explain why it is considered appropriate, as well as its potential pedagogical impact. 
Finally, Novobo will summarize the principles and guidelines for using instructional gestures as expressed by the teachers during the interaction.

We evaluated Novobo with 30 teachers in 10 collaborative sessions. Our goals are twofold: first, to understand how a teachable agent could support teachers in exchanging and building knowledge about instructional gestures; and second, to understand teachers' experiences with Novobo’s core design features to inform future exploration. 
The findings provide rich examples about how teachers externalized and shared their tacit knowledge by formulating collective feedback and engaging in discussion while interacting with Novobo. 
Moreover, the findings surface how teachers built upon each other's experiences and co-constructed understanding of instructional gestures through collaborative demonstration and explanation.
The experiences also demonstrate that teaching an AI mentee together reduced the peer pressure associated with sharing personal and experiential insights.
Based on these empirical findings, we offer implications for future HCI research on leveraging teachable agents to enhance peer learning in teacher professional development.

\section{Related Work}

\subsection{Teachers' Professional Development on Embodied Nonverbal Skills}

In the practice of developing teachers' nonverbal competence, a traditional approach involves the use of video recordings of instructional segments, followed by a review of and reflection on teaching practices \citep{calandra2009using, rich2009video}. By critically analyzing recordings of their own teaching sessions, or those of other more experienced teachers, teachers can refine often overlooked components such as gestures, posture, and eye contact \citep{mccoy2021video}. However, this method demands considerable video analysis and editing efforts, which requires specialized expertise and thus incurs significant time and resource expenses \citep{hamidah2019video}. Currently, the application of AI in the training of teachers’ nonverbal skills remains in its early stages, with only a limited number of studies attempting to classify teachers’ gestures in instructional videos\citep{yoon2024non,pang2023graph,wang2020human,ahuja2019edusense}. However, these methods rely on model training outcomes, making it difficult to analyze the diverse gestures used in teaching, while a single model has limited applicability across different educational contexts and regions. Moreover, these approaches only provide statistical data on gesture types and frequencies, lacking deeper insights and reflections on the reasons behind gesture use in specific teaching scenarios.

An alternative strategy for cultivating nonverbal teaching skills draws upon behavior modification theory \citep{mcdonald1973behavior}. Gestures, as nonverbal skills, often need to be learned through embodied practice \citep{wakefield2021gesture, downey2010practice, kelly2019body}.some studies have ventured into enhancing teachers' nonverbal behaviors through approaches rooted in theoretical knowledge \citep{ishikawa2010can}, which involve understanding the principles behind effective nonverbal communication; discrimination training \citep{wagner1973changing}, where teachers learn to distinguish between effective and ineffective nonverbal cues; or mimicry exercises \citep{feryok2009activity}, where teachers practice imitating model behaviors to improve their own nonverbal communication skills. Building on this, \citet{barmaki2018embodiment} introduced an AI-based gesture training system capable of detecting closed gestures and providing real-time visual or haptic feedback, encouraging users to adopt more open and engaging body language.However, these approaches tend to be overly prescriptive and one-size-fits-all, and rather limited in addressing the context-dependent, experiential nature of teachers' nonverbal behaviors \citep{klinzing1987training}.

The development of knowledge in practice,such as instructional gesturs, always requires attention to, and reflection on, particularly within a contextualized and collective setting, where learning is shaped by the environment and collaboration with others \citep{westerlund2021s}. Thus, leveraging AI to enhance instructional gestures, a highly embodied and collaboration-dependent competence, remains an unexplored research area.

\subsection{Literature Foundations of Gesture-Based Nonverbal Behaviors for Teaching}
In the domain of educational and social sciences, instructional gestures have been explored and validated for their profound impacts in the context of teaching \citep{ goldin2011learning, roth2001gestures, smotrova2017making, novack2015learning, solomon2018applying}. 
For instance, \citet{wakefield2018gesture} demonstrated that gestures can not only enhance learners' attention but also reinforce the teacher's speech. 
Besides, instructional gestures could help students grasp spatial and temporal concepts in second language teaching \citep{matsumoto2017pedagogical}, significantly enhance the comprehension of verbal information, and play an important role in facilitating concept understanding \citep{yoon2024non}. 
A study exploring the significance of instructional gestures from students' perspective revealed that students could interpret the entire intended meaning of teachers' gestures \citep{sime2009because}.
Moreover, the majority of students believed that their teachers' nonverbal behaviors spurred their learning motivation, focused their attention, and imbued them with enthusiasm \citep{marianti2020students}. 
Different types of gestures convey information in various ways. \citet{mcneill1992hand} categorized gestures accompanying speech into Iconic Gestures, Metaphoric Gestures, Deictic Gestures, and Beat Gestures based on their communicative functions. 
\citet{kipp2005gesture} expanded on the classification by dividing gestures into communicative and non-communicative, and further added emblematic gestures—gestures that function independently of speech—to the communicative category. 
Additionally, some researchers have specifically focused on categorizing instructional gestures \citep{lim2019analysing,wei2023students}. 
For instance, \citet{lim2019analysing} classified instructional gestures into Communicative and Performative categories, further dividing Communicative Gestures into Language Correspondent, Language Independent, and Language Dependent types. Nonetheless, McNeill's gesture categorization has been widely based upon by educational researchers as suggested by a large body of research.

Although the positive effects of gesture-based nonverbal behaviors in teaching have been extensively studied, teachers' awareness of these benefits remains insufficient \citep{bambaeeroo2017impact, van2012makes, klinzing1987training}.
Recent literature has called for the inclusion of training in nonverbal behavior, including instructional gestures, in professional development training for teachers \citep{khuman2024impact}. 
Moreover, recent empirical research compared the differences in gesture use between in-service and pre-service teachers and demonstrated the need for increased training in teachers' nonverbal communication skills \citep{yoon2024non}.
These recent results suggest that designing technologies to support the development of teachers' nonverbal behaviors represents a potentially growing area of interest within both educational and Human-Computer Interaction (HCI) research.


\subsection{Teachers' Knowledge Exchange and Co-construction in Collaborative Learning}

The development of teachers' nonverbal behaviors often benefits from collaborative learning experiences shared among peers or with expert educators \citep{seroussi2019reflections}.
Besides, teachers’ pedagogical development through collaborative methods can significantly enhance their ability to integrate training objectives \citep{grasel2010stichwort}. 
Based on the SECI (Socialization, Externalization, Combination, Internalization) model \citep{nonaka1994dynamic}, in such collaborative scenarios there is an opportunity for the exchange and integration of knowledge that is not articulated (tacit) and knowledge that is articulated (explicit), ultimately leading to internalization \citep{westerlund2021s}. 
Additionally, some studies have explored methods to facilitate knowledge construction and learning experiences during collaborative learning \citep{lui2011scripting, falcao2009have, jin2023collaborative}. 

In teacher education programs, Disciplinary Peer Feedback (DPF) is a common collaborative educational approach where teachers critique and provide feedback to each other \citep{seroussi2019reflections}. 
DPF and the discussions it stimulates represent a crucial method of collaborative learning within teacher training. \citet{fleck2009teachers} proposed a wearable camera capturing classroom activities from a teacher’s perspective for quick daily review. Using contextualized image segments, it supported teacher-expert discussions, promoting reflection and suggesting that fragmentary, contextualized materials facilitated low-effort, interactive learning among teachers. In such scenarios, teachers actively partake in the construction of knowledge. They investigate various issues, share perspectives, and foster new insights and knowledge through mutual feedback \citep{hanrahan2001assessing}. 
It is celebrated for its effectiveness in deepening comprehension of the subject matter at hand, offering benefits not only to those being assessed but also to the assessors \citep{dochy1999use}.
For the individuals receiving feedback, peer assessments may shed light on perspectives and knowledge not covered in instructor evaluations \citep{seroussi2019reflections} and encourage reflective thinking \citep{klimenko2015formative}. 
For the assessors, the act of delivering feedback strengthens their subject matter understanding, as it requires pinpointing and articulating the essential components within the material learned \citep{cartney2014exploring}. In this sense, DPF mirrors the ``learning by teaching'' approach, as individuals critically evaluate peers' work and articulate their understanding, thereby reinforcing their own knowledge through the process of teaching or giving feedback to others \citep{Topping1998}. 

Yet, despite the advantages of DPF, teachers may exhibit  reluctance towards peer evaluation, largely owing to social dynamics \citep{seroussi2019reflections}.
The aversion to the potentially uncomfortable role of offering criticism has been identified as a cause for inconsistencies in the reliability of peer assessments \citep{brown2012assessing}. 
Our study aims to explore how to harness the benefits of DPF and, therefore, ``learning by teaching'', by investigating the role teachable AI agents can play in enhancing social dynamics and knowledge exchange in teachers' peer learning while minimizing feelings of criticism.



\subsection{Teachable Agents}
Learning by Teaching (LBT) is a pedagogical approach wherein learners articulate and restructure their existing knowledge \citep{jin2023teach}.
It is reported to be effective in constructing meaning \citep{choi2021reconsidering}.
Teachable agents (TAs) are developed based on the principle of LBT, actively engaging learners by having them instruct the TA \citep{song2017designing}. 
Research suggests that when incorporating AI tools in education, reflective practices should be integrated \citep{mogavi2024chatgpt}, and teachable agents have been shown to hold potential in supporting such practices \citep{roscoe2007understanding}. For example,\citet{chhibber2022teachable} designed the process of data classification by crowd workers as a way to teach a teachable agent, and found that this approach helps establish trust between humans and AI. 

With the advent of LLMs, researchers \citep{jin2023teach, ma2024teach} explore leveraging their powerful code generation capabilities to design TAs that facilitate LBT in programming education.
TAs have demonstrated encouraging outcomes in enhancing students' performance, fostering their self-explanation skills, and increasing their receptiveness to constructive feedback \citep{silvervarg2021teachable, matsuda2011learning, chase2009teachable, davis2003intelligent}.
Furthermore, it has been explored that when students collaboratively teach an agent, it promotes discussion among them, ultimately leading to the integration of different viewpoints \citep{yu2009teachable}.


To the best of our knowledge, current teachable agents are almost exclusively designed for students. 
However, research indicates that if teachers assume a more central role as active learners in their own professional development, nurturing their growth, they are more likely to engage in responsible and productive inquiries into their practice \citep{cochran1993inside, aschermann2015collaborative}.
Moreover, instructional gestures, which are often hard to notice and reflect on, can potentially be better developed and facilitated reflective knowledge-building through LBT \citep{roscoe2007understanding}.
The above suggests a timely opportunity to explore teachable agent systems that can support teachers in fostering their development and enhancing their pedagogical practice, particularly in learning instructional gestures.
 


\subsection{Research Questions and Contribution to HCI}

In sum, despite growing recognition of the importance of developing instructional gestures in teaching, there has been limited exploration of how teachable agent systems can support teachers in this area. This presents a timely opportunity to investigate how such systems can foster teachers' development and enhance their pedagogical practice in learning and refining instructional gestures.
Addressing this gap in the literature, recent advancements in generative AI (GenAI) and LLMs offer the potential to develop a neutral agent acting as a mentee that teachers can instruct. This agent could assist teachers in facilitating the collective externalization and reflection on their instructional behaviors, particularly nonverbal gestures, within a structured environment that promotes peer learning and professional growth.

The contribution of this paper is the design and evaluation of Novobo, a teachable AI agent that acts as a mentee to stimulate peer learning among teachers focused on instructional gestures. By using Novobo both as a novel design and as a research tool, we aim to answer the following questions. 

The first question seeks to gather empirical evidence on the role a teachable AI agent can play in facilitating group discussions and the development of instructional gesture strategies among teachers. Specifically, it explores how the agent can support collaborative learning and enhance teachers' nonverbal instructional skills. The first research question is:

\textbf{RQ1:} To what extent, and in what ways, can a teachable AI agent support teachers' peer learning of instructional gestures?

The second question focuses on eliciting teachers' reactions to the specific design features of Novobo, with the goal of evaluating the teachable agent's design and identifying potential design implications for teacher-focused teachable AI agents. The research question is:

\textbf{RQ2:} How do teachers experience and respond to the core design features of Novobo?

To investigate these questions, we conducted an empirical study involving 30 teachers in 10 collaborative sessions. 
Rich empirical findings depict how teachers externalized and socialized their tacit knowledge through both verbal and bodily expressions, and how they build new understanding together through collaboratively teaching Novobo. 
Moreover, teachers' experiences about interacting with Novobo also contextualized how certain design features might influenced their collaboration and knowledge sharing. This work thereby makes a towfold contribution to HCI:

\begin{itemize}

    \item \textbf{We designed and implemented Novobo}, a teachable AI mentee capable of discussing instructional gestures with traceable knowledge source through Retrieval-Augmented Generation and multi-agent collaboration, and requesting teachers verbal feedback and bodily demonstration to stimulate their knowledge exchange. 
    \item \textbf{Results from an empirical research study (N=30)} of how teachers exchanged and co-constructed embodied knowledge related to instructional gestures with the support of Novobo.
    
\end{itemize}

\section{Design}
This section describes the key activities we undertook during the design phase, which led to the final Novobo system, including: (1) knowledge preparation, where we compiled the necessary domain knowledge to ensure that teachers could trust the gestures generated by the teachable agent, thereby facilitating effective discussions about instructional gestures., and (2) collaborative design, involving 6 teachers to identify user needs, refine the initial design concept, and gather valuable formative feedback from a user-centric perspective.

\subsection{Knowledge Preparation}

Although teachable agents acts as mentees instead of mentors, prior research highlights the importance of knowledge preparation for these agents \citep{brophy1999teachable}. The design of teachable agents should integrate relevant learning resources, including reliable domain knowledge, to ensure the educational value of generated content \citep{blair2007pedagogical}. Additionally, providing teachers with up-to-date theoretical knowledge on instructional gestures has been shown to enrich their teaching practices \citep{thompson2014teachers}. 
To combine both theoretical knowledge with reliable, traceable sources and practical insights from practitioners, our knowledge preparation focused on two key areas: (1) organizing theoretical knowledge on instructional gestures, and (2) inviting teachers to annotate examples of gesture descriptions.

\begin{table}
  \caption{Demographic and professional information of teachers in the user study.}
  \label{tab:gesture_types}
  \small
  \begin{tabular}{@{}p{0.2\textwidth}p{0.8\textwidth}@{}}
    \toprule
    \textbf{Gesture Type} & \textbf{Definition} \\
    \midrule
    Iconic gesture & Iconic gestures are hand and body movements that visually represent the characteristics or features of concrete objects, actions, or entities \citep{mcneill1992hand,lim2019analysing}. \\ 
    \midrule
    Metaphoric gesture & Metaphoric gestures symbolize abstract ideas or concepts, not directly portraying physical objects but representing them through symbolic hand movements \citep{mcneill1992hand,lim2019analysing}. \\ 
    \midrule
    Deictic gesture & Deictic gestures are used to draw attention to an object, location, or person, functioning primarily to indicate or "point out" specific spatial information, corresponding to the person, object, or thing mentioned by the teacher \citep{mcneill1992hand,lim2019analysing}. \\ 
    \midrule
    Emblematic gesture & Emblematic gestures are conventional gestures that have a specific, culturally understood meaning, such as a thumbs-up for approval, requiring no verbal accompaniment to convey their message \citep{mcneill1992hand,kipp2005gesture,lim2019analysing}. \\ 
  \bottomrule
\end{tabular}
\end{table}

\subsubsection{Theoretical Knowledge Related to Instructional Gestures}
Based on related literature reviewed in the previous section, we organized the theoretical knowledge concerning the types of gestures and the instructional intentions conveyed by gestures.
In terms of gesture types, we referenced several existing classifications \citep{mcneill1992hand,ec2000types,lim2019analysing,kipp2005gesture} and based our approach on McNeill’s co-speech gesture taxonomy \citep{mcneill1992hand} , which has been widely adopted in instructional gesture research. However, we excluded beat gestures because they primarily involve subconscious rhythmic movements and do not effectively convey clear intentions or specific content. Additionally, we included emblematic gestures \citep{kipp2005gesture}, which function independently of speech and frequently used in teaching (e.g., a thumbs-up). Consequently, we organized the topological knowledge regarding instructional gestures in four major categories: iconic, metaphoric, deictic, and emblematic, as summarized in Table \ref{tab:gesture_types}.
In terms of instructional intentions, we synthesized existing research \citep{kaneko2015reflective,alibali2013students,alibali2014teachers,wakefield2018gesture,kartchava2020investigating,bosmans2022teachers} and identified four key intentions that teachers use gestures to convey: attracting students’ attention, imparting new knowledge, explaining complex concepts, and providing positive feedback. We summarized these two aspects with types, definitions, examples, and supporting literature, and validated and finalized them with teachers in the collaborative phase described below.

\subsubsection{Gesture Examples Gathered from Teacher Annotators}

To enrich the teachable agent’s with teachers' practice-based knowledge and ensure the quality of its output description of instructional gestures,  we also constructed a gesture example set. We collected 117 clips from classroom teaching videos of five primary school teachers, covering three different subjects, sourced from an online open course platform. These clips were selected to ensure they represented the gesture types and instructional intentions summarized during the knowledge preparation phase. We then invited seven teachers, each from a different primary or middle school, to annotate these clips, with the annotations including detailed description of the gestures and explanations of the instructional intentions behind them. Each teacher annotated approximately 15 data points and was compensated \$10 dollars. These annotated examples, along with the organized theoretical knowledge, form the knowledge base for the teachable agent's retrieval augmented generation (RAG).

\subsection{Collaborative Iteration of the Initial Design with Teachers}
\label{formative}

In this phase, we collaborated with 6 school teachers through workshops, to iterate concepts and concertize teachers' needs and desirable interaction qualities.
Based on related research and our design motivations, we established a preliminary design for a teachable agent used for teachers' study groups that could generate ideas about instructional gestures in response to teachers' input teaching scenarios leveraging RAG. Given that instructional gestures, as tacit knowledge, require embodied practice \citep{downey2010practice, kelly2010two, wei2024leveraging}, we integrated a webcam ``mirror'' feature for teachers to record and review their gestures for reflection. We then leveraged our early prototype and mockups as a technology probe for collaborative design iterations with teachers, helping identify user needs, test early-stage technology, and inspire potential technology integration \citep{hutchinson2003technology}.

\subsubsection{Technology probe}

Our technology probe (TP) included several static webpages with a teachable agent providing\textit{ pre-defined responses}. We collected 30 teaching scenarios from classroom videos available online, covering summarized instructional intentions and gesture types. Gesture descriptions were pre-generated using LLM and RAG, displayed when teachers clicked on a scenario. Teachers can provide comments on the generated gesture, then record their own gestures via webcam, describe and explain them using a text box. After that, all system-generated intermediate data (including gesture types, instructional intentions, and references) were presented to teachers for further feedback. 

\subsubsection{Procedure}

We recruited 6 teachers (all self identified females, teaching three grades and two subjects) to participate in two rounds of design workshops, with two teachers in the first round and four in the second. Each teacher received a 15 dollar compensation. Each workshop consisted of a prototype experience session followed by a semi-structured interview. We analyzed the video recordings of the workshops using affinity diagram to formulate insights. For clarity, we refer to the teachers as T1 to T6.

\begin{table}
  \caption{Instructional intentions conveyed by gestures and their descriptions.}
  \small
  \label{tab:instrcutional intentions}
  \begin{tabular}{@{}p{0.2\textwidth}p{0.8\textwidth}@{}}
    \toprule
    \textbf{Instructional Intention} & \textbf{Description} \\
    \midrule
    Explaining complex teaching content & Teachers use gestures when teaching complex content can significantly help students’ understanding, especially in micro- and macro-levels or processes that are difficult to visualize\citep{kang2013different,ahmadi2023classification,alibali2014teachers,abakumova2021non}. \\
    \midrule
    Attracting students' attention & Teachers use gestures to capture students' attention, focusing them on the target concept or project\citep{alibali2014teachers,bosmans2022teachers,kartchava2020investigating} .\\
    \midrule
    Providing positive feedback to students & Teachers enhance positive feedback through gestures, improving students' sense of competence, and promoting teacher-student relationships, motivation, and engagement \citep{ahmadi2023classification,bergold2023teacher,shin2023pedagogical,yoon2024non}. \\
    \midrule
    Imparting new knowledge & Teachers enhance the memorability of knowledge through gestures when imparting new knowledge\citep{kang2013different,wakefield2018gesture,alibali2014teachers}.\\
    \midrule
     Role modeling & Teachers model the gestures and actions that students, especially in the lower grades, should follow and use in the classroom, so that students can imitate and acquire them.(Added by Teachers in Practitioner Involvement)\\
  \bottomrule
\end{tabular}
\end{table}

\subsubsection{Insights} 
Teachers' input crucially helped validate and evolve our design concepts. For instance, drawing from their practical knowledge, they not only confirmed and contextualized the existing categories of gestures (see Table \ref{tab:gesture_types}) and instructional intentions (see Table \ref{tab:instrcutional intentions}) based on literature , but also pinpointed a new type of intention not covered by the literature. Namely, they highlighted the use of ``role modeling,'' where teachers demonstrate gestures as examples of desired student behavior. As T2 noted, \textit{``We often use gestures like raising hands or placing hands properly on the desk to help younger students develop behavioral norms.''} Following their advice, ``role modeling'' was included into the instructional intentions (see Table \ref{tab:instrcutional intentions}).
Moreover, teachers stressed that gestures are context-dependent, varying by grade level, subject, topic, and personal teaching style. This confirmed that a prescriptive, one-size-fits-all approach would be ineffective. Instead, the TP proposed gestures were more valuable as prompts for teachers' reflection and experience-sharing, aligning with our intended design. 
To enhance this, in the initial interaction stage where teachers input or select teaching scenarios for the TP, we enriched the options for scenarios with subject areas, grade levels, and lesson topics, encouraging more thoughtful discussions about the contextual aspects of instructional gestures.

Aside from these concrete adjustments, three key insights emerged from the teachers' feedback (see Table \ref{table:requirements}), shaping the core design features (\textbf{DF1-3}) of our final design outcome:

\textbf{DF1: Supporting learning through well-referenced domain knowledge.}
Initially, the domain knowledge base was solely meant to afford Retrieval-Augmented Generation and to serve as a prompt to ensure the reliability and educational meanings of teachable agent's pre-defined responses. Whereas, teachers preferred the TP to explicitly reference supporting literature when explaining its suggested gestures.They noted that this approach not only helps them access cutting-edge knowledge but also reduces their skepticism toward AI- generated responses.Therefore, this has become a core feature in our final design.

\textbf{DF2: Supporting learning through embodied practice with a skeletal mirror.}
We observed that teachers naturally performed gestures to aid their reflection in practice when interacting with the TP. They confirmed that the webcam's mirror view supported their self-observation during embodied practice and helped them focus on their movements. However, some teachers felt nervous and self-conscious seeing their reflection. As suggested by the teachers, we decided to display only the skeletal structure of their bodies, excluding facial features and details.

\textbf{DF3: Supporting learning through mentor-mentee dynamics between teachers and AI.}
We found that emphasizing the mentor-mentee relationship between teachers and the TP helped clarify the TP’s intended use. Initially, some teachers viewed the TP’s responses as recommendations, not realizing they were meant to teach it. 
Teachers pointed that the mentor-mentee model is a common training method across schools, where a novice teacher is paired with a more experienced mentor. This concept motivated framing the TP as a novice AI teacher skilled in theory but lacking practical experience.Teachers can be their mentors, guiding the TP on instructional gestures, while the TP connects scenarios with relevant literature. This narrative not only helped teachers better understand their role in interacting with the teachable AI agent but was also confirmed by users to reduce resistance and foster a more inclusive learning environment.

\begin{table}
  \caption{Requirements and exemplar quotes based on teacher feedback}
  \small
  \label{table:requirements}
  \begin{tabular}{@{}p{0.35\textwidth}p{0.6\textwidth}@{}}
    \toprule
    \textbf{Requirement} & \textbf{Exemplar Quotes} \\
    \midrule
    Facilitate knowledge acquisition and build trust in AI.
    & ``I want to learn some cutting-edge theories about instructional gestures.'' (T1) \newline 
    ``Seeing the source of knowledge lets me know it (AI) isn’t just making things up.'' (T4) \\
    
    \midrule
    Enable comfortable observation and refinement of instructional gestures to promote self-reflection and improvement. 
    & ``Seeing myself in the mirror makes me feel too self-conscious.'' (T1) \newline 
    ``When recording, I want to show my best self.'' (T2) \\
    \midrule
    Explain the relationship between teachers and AI in a way that is easier to understand. 
    & ``I thought it was here to teach me.'' (T1) \newline 
    ``My mentor taught this (using instructional gestures) early in my career.'' (T5) \\
    \bottomrule
  \end{tabular}
\end{table}
 
\subsection{Novobo: System Design and Implementation}
\label{Novobo}

This section describes the final design and implementation of Novobo, a teachable agent that embodies the three core design features (\textbf{DF1-3}) and supports teachers' peer learning of instructional gestures.

\subsubsection{User Interface Design and Usage Flow}
\label{stages}

Novobo presents itself as an AI novice teacher, seeking to learn from its mentors---the teachers---about the use of instructional gestures in teaching (\textbf{DF3}). While Novobo has theoretical knowledge, it lacks practical experience. It encourages teachers to pose teaching scenarios, similar to a mentor guiding a mentee. In response, Novobo generates gesture descriptions based on theory, explaining why it thinks these gestures may align with the instructional intentions of the scenario. Teachers then provide feedback on the suitability of the suggestions and demonstrate gestures they find appropriate. Through this interaction, Novobo aims to foster peer learning by encouraging knowledge exchange and co-construction among teachers.

Resulted from our design iteration, Novobo's interaction flow was finalized into four stages: question posing, commentary, demonstration, and explanation (see Figure \ref{fig:interface}). These stages form a repeatable cycle that users can follow as many times as desired. Conversations can be done through both text input and a speech-to-text function for users' convenience. These stages are detailed as follows:

\begin{figure*}[h]
    \centering
    \includegraphics[width=1\linewidth]{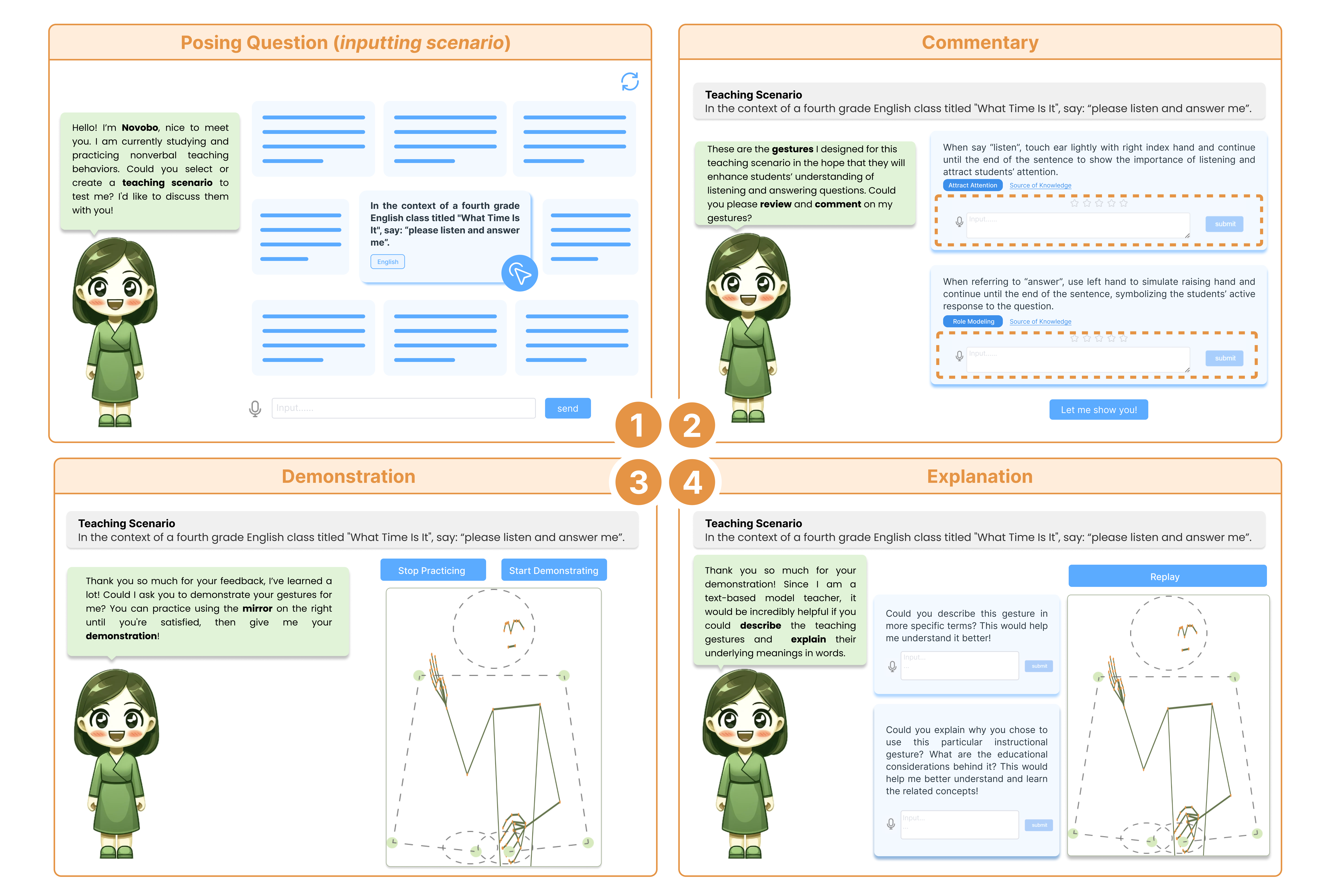}
    \caption{An overview of the Novobo interface. (1) \textit{Question Posing (inputting scenario)}: A teaching scenario is provided to Novobo as the posed question; (2) \textit{Commentary}: Novobo invites users to rate and comment on the instructional gestures it generates; (3) \textit{Demonstration}: Users practice and demonstrate gestures using the skeletal mirror; (4) \textit{Explanation}: Users describe and explain their own instructional gestures.}
    \label{fig:interface}
\end{figure*}

\textbf{Posing Question.} This stage involves inputting scenarios and viewing Novobo's answers. (1) Inputting scenarios: teachers can select the provided teaching scenarios from the interface or input custom scenarios by typing or using voice input (Figure \ref{fig:interface}-\circled{1}). (2) Viewing answers: Novobo then generates and displays several gesture descriptions for teachers to review. For each gesture, it provides the corresponding instructional intention, with relevant theoretical knowledge available through foldable sections \textbf{(DF1)}.

\textbf{Commentary.} After reviewing the answers, teachers need to rate each gesture generated by Novobo by giving stars and also provide comments to justify their ratings, noting strengths, weaknesses, and offering suggestions for improvement \textbf{(DF3)}. After rating and commenting on each gesture, they need to click the submit button, and Novobo will response what it has learned based on the teachers' feedback (Figure \ref{fig:interface}-\circled{2}).

\textbf{Demonstration.} Novobo prompts teachers to demonstrate a gesture they find appropriate for the given teaching scenario. This is done using a skeletal mirror view on the interface, which shows the teachers' bodily movements as an abstract skeletal representation (\textbf{DF2}). Teachers can practice their gestures by pressing the practice button. When satisfied, they can press the record button to begin. A three-second countdown precedes the recording. Upon completion, the recorded skeletal movement is played back for review (Figure \ref{fig:interface}-\circled{3}).

\textbf{Explanation.} In the final stage, teachers are asked to provide a detailed description of their demonstrated gesture and explain their reasoning for using it in the given teaching scenario. Novobo then synthesizes all the information from the interaction, summarizing the knowledge shared by the teacher and acknowledging the insights gained from their input (Figure \ref{fig:interface}-\circled{4}).

\begin{figure*}[t]
    \centering
    \includegraphics[width=1\linewidth]{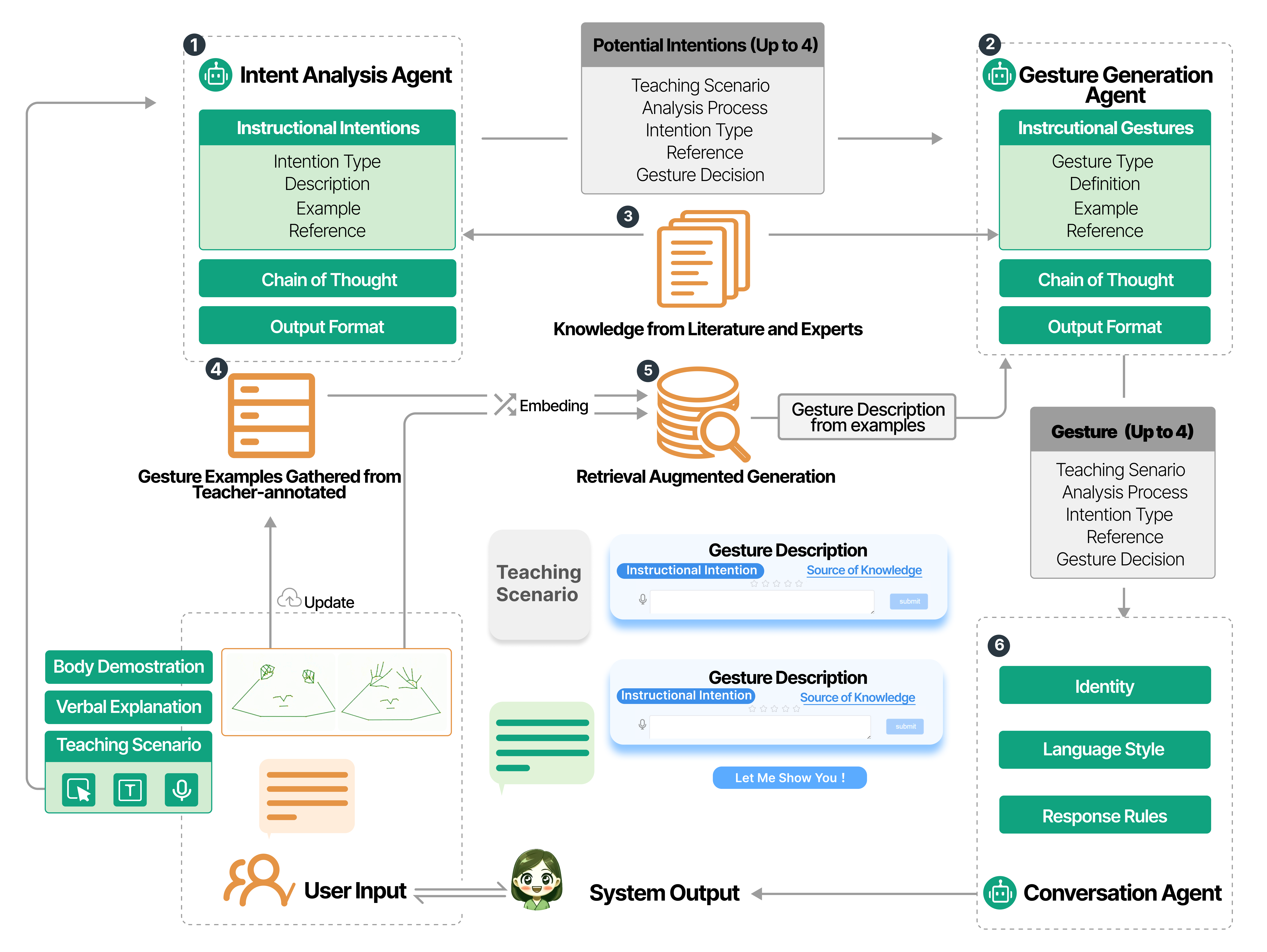}
    \caption{ The multi-agent architecture and workflow of Novobo}
    \label{fig:system}
\end{figure*}

\subsubsection{Multi-agent Architecture and Workflow of Novobo}

To enable Novobo to generate diverse, knowledge-based responses and facilitate real-time conversations, we chose to leverage the ability of LLMs.
This section details how we utilize a multi-agent system based on LLMs to realize the workflow for the generation of gesture descriptions and the conversations between Novobo and the user.
We introduce this workflow in two parts: (1) how we integrate knowledge into the LLM workflow using Retrieval Augmented Generation (RAG) and prompt engineering, and (2) how we realized the gesture description generation and conversation via the multi-agent workflow. The figure \ref{fig:system} illustrates an overview of the architecture and workflow of the system.

\textbf{RAG and Prompt Engineering.} 
As mentioned in our knowledge preparation, we have established a domain knowledge base including the theoretically grounded categories of gesture types and instructional intentions (see Table \ref{tab:gesture_types} and Table \ref{tab:instrcutional intentions}), as well as gesture examples gathered from teacher annotators. 
This knowledge base has Incorporated in our RAG module and prompt engineering.
We encoded theoretical knowledge (see Figure \ref{fig:system}-\circled{3}) into the LLM's system prompt \citep{si2022prompting} and enhanced its reasoning abilities through chain of thought \citep{wei2022chain,yao2024tree}. The LLM analysis instructional intentions and gesture types based on  theoretical knowledge. 
To enhance the quality of gesture generation, we used the gesture examples (Figure \ref{fig:system}-\circled{4}) as a retrieval source, applying RAG (see Figure \ref{fig:system}-\circled{5}) \citep{lewis2020retrieval,gao2023retrieval} to retrieve the most contextually similar gesture description, which was then incorporated into the prompt as a few-shot example \citep{brown2020language}. All of our system prompts are provided in the supplementary materials.

\textbf{Multi-Agent Collaboration.} To avoid degrading model performance with lengthy prompts, we utilized two agents \citep{wu2022ai}: one to analyse instructional intentions of the teaching scenario and decide whether to generate a gesture and the other to determine what gesture to generate. 
The intention analysis agent (see Figure \ref{fig:system}-\circled{1}) receives the teaching scenario from the user, generates multiple instructional intentions, and determines whether each intention requires a gesture. 
The gesture generation agent (see Figure \ref{fig:system}-\circled{2}) then determines the gesture type and generates the corresponding gesture description based on the examples retrieved via RAG. We limit each scenario to a maximum of four gestures.
We implemented another conversation agent (see Figure \ref{fig:system}-\circled{6}) to manage the dialogue between Novobo and the user \citep{wei2024leveraging,seo2024chacha}. 
This agent continuously processes user input and the generated content at each stage. 
We identified its identity, language style, and response rules through prompt engineering to adopt an mentee role for the user. Additionally, it guides the user’s actions at each stage through language prompts and briefly analyzes and summarizes the user’s inputs.

\subsubsection{Implementation}
The Novobo application and RAG are implemented in Python. The multi-agent system utilizes OpenAI's pre-trained LLMs. Specifically, the intention analysis and gesture generation agents employ the more analytically capable GPT-4-turbo model, while the conversation agent uses GPT-3.5-turbo to improve response speed. RAG is powered by OpenAI's API, utilizing the text-embedding-3-small model for efficient and accurate text embedding. The server is built with Flask. The interface is developed in JavaScript using Vue.js. It communicates with the server by exchanging messages between the user and the agent hosted on the server.

\section{User Study and Evaluation}

We evaluated the Novobo system with 30 teachers in 10 collaborative sessions to examine: i) to what extent and in what ways the designed teachable AI agent supported teachers' peer learning of instructional gestures (\textbf{RQ1}), and ii) how teachers experienced and responded to Novobo's core design features (\textbf{RQ2}).

\subsection{Participants}

We recruited 30 teachers (including 18 self-identified females, with an average age of 31.8 years, and an average of 8 years of teaching experience) from 7 local primary (including two special education schools, as research shows gestures aid inclusive education\citep{vogt2017some}) schools from [Anonymized for review] through social media platforms, with detailed demographic and professional experience details provided in the table \ref{demographic}. To gather insights from a range of teaching subjects, teachers from diverse disciplines were included---spanning math, art, science, language, and more---as detailed in the table \ref{demographic}. It is common in local primary schools for teachers of different subjects to engage in group discussions to share knowledge and teaching strategies. As a result, and to address our research question focused on peer learning, participants  were organized into 10 groups for the user study, including two groups of 6 teachers from special education schools. To reflect their existing professional development context, each group consisted of teachers who were already peers from the same school and participated in regular study group discussions. Each group included from two to four peers. Evaluation sessions lasted 90 minutes (see details below), and each teacher was compensated \$15. For clarity, we labeled the groups as G1 to G10 and the participants as P1 to P30.

\begin{table*}[tb]
\centering
\caption{Demographic and professional information of teachers in the user study.}
\vspace{-3mm}
\label{demographic}
\small
\resizebox{1\linewidth}{!}{
\begin{tabular}{ccccccc}
\toprule
\textbf{Group} & \textbf{ID} & \textbf{Gender (Self-Identity)} & \textbf{Age} & \textbf{Experience (Years)} & \textbf{Subject} & \textbf{School Type} \\
\midrule
G1  & P1  & F  & 30s  & 2   & Art                & Regular School \\
    & P2  & M  & 50s  & 34  & Math               & Regular School \\
    & P3  & M  & 20s  & 3   & Art                & Regular School \\
    & P4  & F  & 40s  & 26  & Language           & Regular School \\
G2  & P5  & F  & 30s  & 8   & Information Science & Regular School \\
    & P6  & F  & 40s  & 26  & Information Science & Regular School \\
    & P7  & M  & 30s  & 12  & Information Science & Regular School \\
G3  & P8  & M  & 20s  & 1   & Information Science & Regular School \\
    & P9  & M  & 40s  & 24  & Information Science & Regular School \\
G4  & P10 & M  & 20s  & 4   & Language           & Regular School \\
    & P11 & F  & 20s  & 1   & Language           & Regular School \\
    & P12 & F  & 20s  & 1   & Math               & Regular School \\
G5  & P13 & F  & 20s  & 4   & Language           & Regular School \\
    & P14 & F  & 20s  & 1   & Math               & Regular School \\
    & P15 & M  & 20s  & 1   & Language           & Regular School \\
    & P16 & F  & 20s  & 1   & Science            & Regular School \\
G6  & P17 & M  & 20s  & 1   & Labor Studies      & Regular School \\
    & P18 & F  & 20s  & 5   & Math               & Regular School \\
    & P19 & F  & 30s  & 4   & Science            & Regular School \\
G7  & P20 & M  & 20s  & 1   & Physical Education and Health & Regular School \\
    & P21 & F  & 20s  & 2   & Language           & Regular School \\
    & P22 & M  & 40s  & 17  & Information Science & Regular School \\
G8  & P23 & F  & 40s  & 18  & Math               & Regular School \\
    & P24 & F  & 40s  & 23  & Math               & Regular School \\
G9  & P25 & F  & 20s  & 3   & Information Science \& Life Skill & Special Education School \\
    & P26 & F  & 30s  & 7   & Language \& Life Skill & Special Education School \\
    & P27 & F  & 20s  & 1   & Math \& Life Skill    & Special Education School \\
G10 & P28 & M  & 20s  & 3   & Drawing and Crafts & Special Education School \\
    & P29 & F  & 20s  & 1   & Drawing and Crafts & Special Education School \\
    & P30 & M  & 30s  & 8   & Physical Education and Health & Special Education School \\
\bottomrule
\end{tabular}}
\end{table*}

\subsection{Procedure}
Each group session included the following activities: 

\textbf{Introduction:} Researchers introduced the background and planning of the evaluation session, followed by the demonstration of the system and its interaction flow. The introduction lasted about 15 minutes.

\textbf{Interaction (RQ1):} Each group of teachers was invited to interact with Novobo together, collaboratively engaging in the four-step interaction flow of problem posing, commentary, demonstration, and explanation. This interaction activity lasted approximately 45 minutes.

\textbf{Focus group discussion (RQ2):} Following the system interaction, we conducted a focus group aimed at discussing teachers' perceptions of Novobo's interaction and their experiences with its core design features. This discussion lasted around 30 minutes.

\subsection{Analysis}

All sessions were recorded, capturing both video and audio data. We collected approximately 450 minutes of process data and 300 minutes of data from the 10 focus groups.
To explore how Novobo supported teachers’ peer learning of instructional gestures (\textbf{RQ1}), we conducted vignette analysis \citep{atzmuller2010experimental} using video recordings of teachers' interactions with Novobo. Two researchers collaborated to identify vignettes from the four stages of interaction that demonstrated knowledge exchange in peer learning, guided by relevant theories \citep{nonaka1994dynamic}, which explains the process of Socialization, Externalization, Combination, and Internalization (SECI) of knowledge within a group (see Figure \ref{fig:esc}). An exemplar vignette and a set of representative examples were curated to characterize teachers' peer learning at each interaction stage with Novobo. 

To understand how teachers experienced and responded to Novobo's core design features (\textbf{RQ2}), we employed thematic analysis \citep{clarke2017thematic} on the transcriptions of the semi-structured interview data. Two researchers collaboratively coded the transcripts, iteratively refining themes \citep{mcdonald2019reliability}. A codebook was initially developed based on the research question and further refined through regular meetings. Using the finalized codebook, the researchers conducted collaborative coding to formulate the final thematic structure.

\section{Results for RQ1: Teachers' Knowledge Exchange and Co-Construction Supported by Novobo}

\begin{figure*}[h]
    \centering
    \includegraphics[width=1 \linewidth]{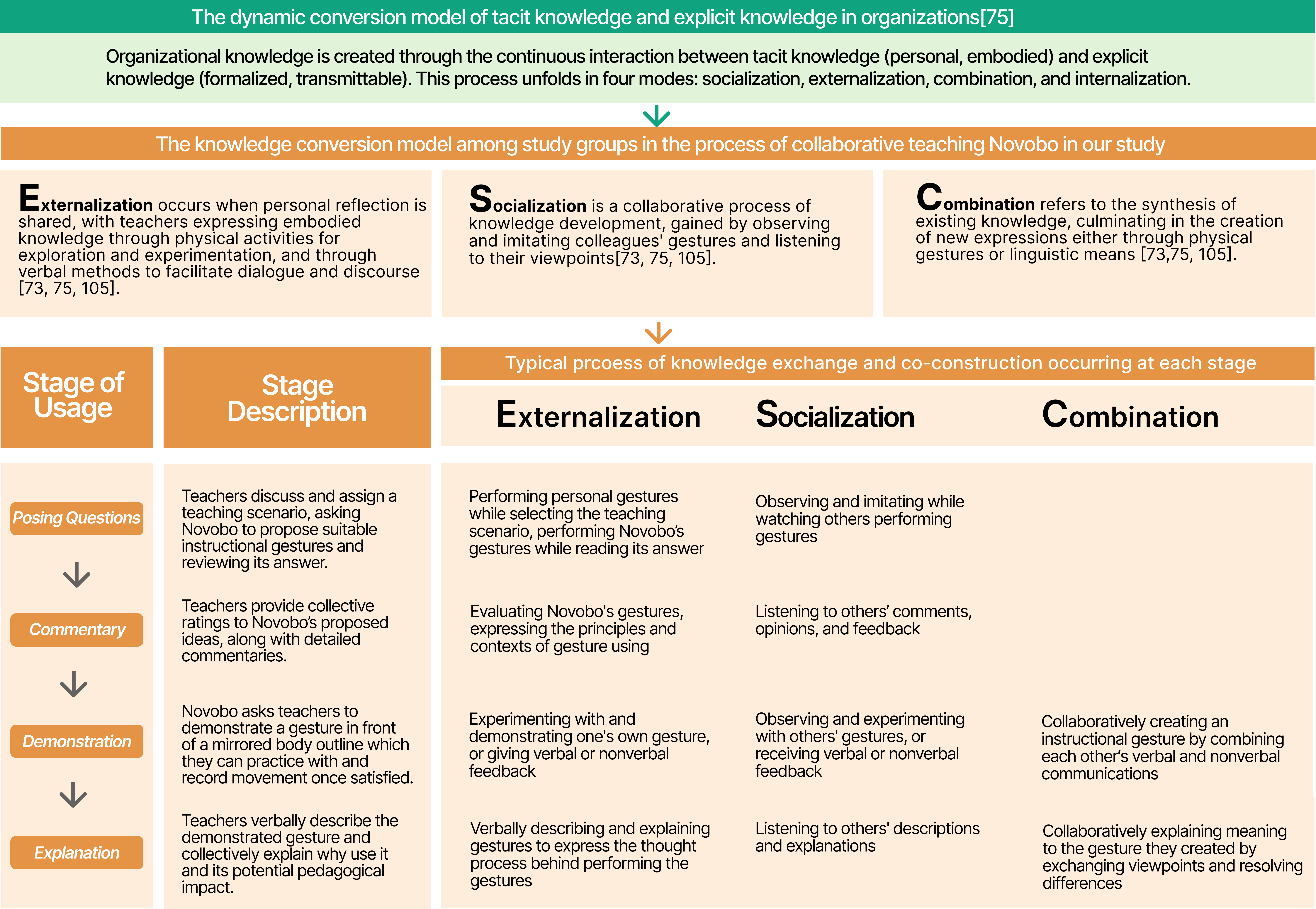}
    \caption{The description of knowledge externalization, socialization, and combination that occur when teacher study groups collaboratively teach Novobo, as well as the typical processes of externalization, socialization, and internalization that take place during the four stages of \textit{Posing Questions}, \textit{Commentary}, \textit{Demonstration}, and \textit{Explanation}.
}
    \label{fig:esc}
\end{figure*}

In this section, we present the findings for \textbf{RQ1}, organized according to the four stages of interaction with Novobo—namely, \textit{Question Posing}, \textit{Commentary}, \textit{Demonstration}, and \textit{Explanation} (described in Section \ref{stages}). We illustrate how user groups were supported by the teachable agent, Novobo, in facilitating group discussions and promoting collaborative learning on instructional gesture knowledge throughout these stages. We employ the SECI framework \citep{nonaka1994dynamic, westerlund2021s, mendoza2022assessing}, which explains the processes of externalization (\textbf{E}), socialization (\textbf{S}), and combination (\textbf{C}) as a theoretical lens to describe knowledge exchange and co-construction within the context of teachers' collaboratively embodied learning of instructional gestures (see Figure \ref{fig:esc}).

\begin{figure*}[h]
    \centering
    \includegraphics[width=1\linewidth]{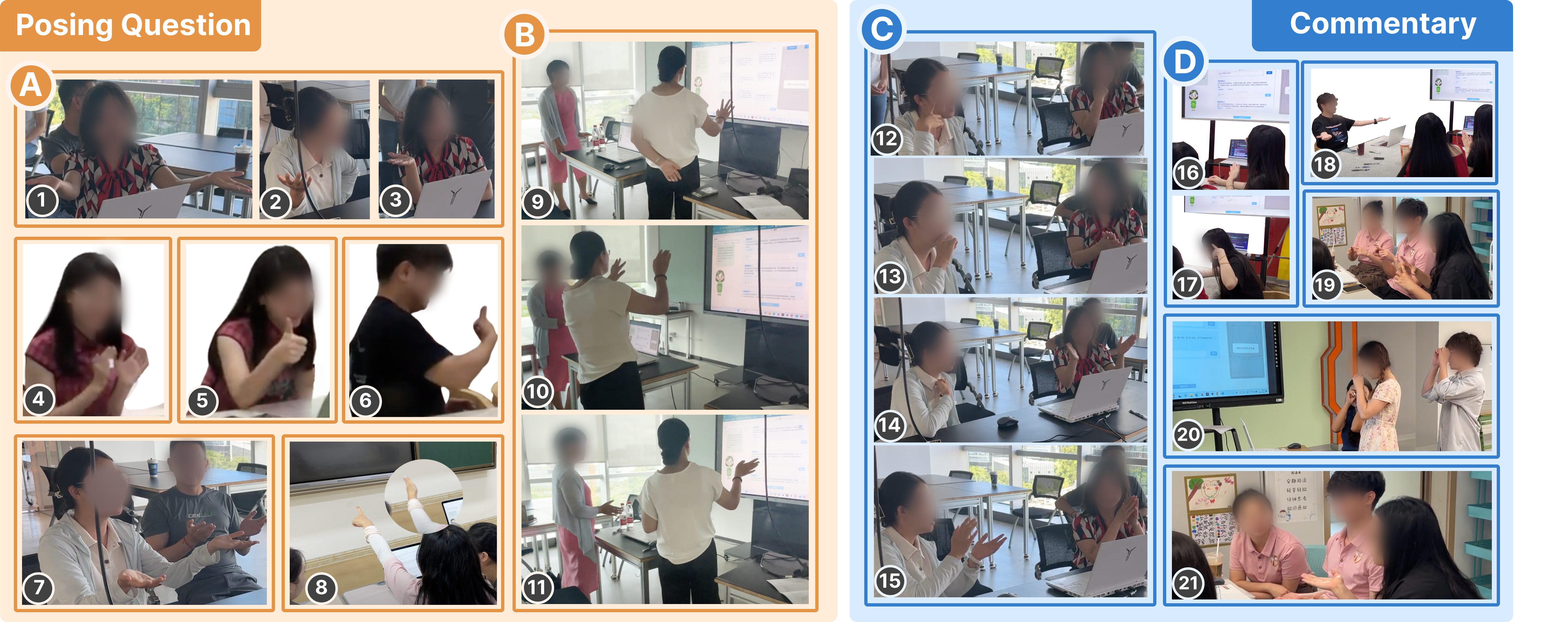}
    \caption{Typical example scenarios of users engaging in the \textit{Posing Question} and \textit{Commentary} stages of the system.}
    \label{fig:posing_and_commen}
\end{figure*}

\subsection{Posing Question --- Awareness and Reflection}

During this phase, teachers collectively select or input a teaching scenario to Novobo. 
Based on the chosen scenario, Novobo generates corresponding instructional gestures, which are then presented to the teachers in textual form. 
This presentation includes the description of the instructional gestures, the instructional intent and the sources of the related knowledge.

\textbf{Vignette.} Figure \ref{fig:posing_and_commen}-\circled{B} provides a typical example of teachers' collaboration at this stage. 
The teachers from Group 8 collectively selected a scenario for depicting ``The leaves gently fell down.''
As G8P24 read the scenario, she simultaneously performed a gesture moving her right hand downward with the palm facing down (\textbf{E}). 
At that moment, G8P23 were observing her actions (\textbf{S}).
When Novobo offered its response: ``\textit{gently waving both hands, imitating leaves fluttering in the air until the end of the sentence,}'' 
G8P24 first mimicked this gesture and asked G8P23, ``Will we do this in class?'', 
G8P23 then began to imitate the gesture herself. 
After several attempts, she was not entirely satisfied and chose to simulate the falling motion using only her right hand (\textbf{E}). 
After observation and physical engagement (\textbf{S}), G8P24 also agreed with G8P23's instructional gesture.

\textbf{Knowledge exchange (E\&S).} 
As depicted in Figure \ref{fig:esc}, during this stage, teachers primarily externalized knowledge through bodily movements (\textbf{E}) and socialized by observing the gestures of other teachers (\textbf{S}). 
On the one hand, when reading through a scenario, teachers mentally immersed themselves in the teaching context and perform non-verbal behaviors that they would use in such situations (see Figure \ref{fig:posing_and_commen}-\circled{1}, \circled{2}, \circled{3}, \circled{5} and \circled{6}). 
Through this externalization, some teachers noticed differences in their gestures compared to those of their colleagues. For instance, in Group 2, when the reading reached ``What else is there?'', G2P6 first demonstrated a gesture and then sought feedback from peers on her approach. Upon realizing that her colleague G2P5's gesture differed from hers, she began mimicking G2P5's gesture and explored the reasons behind this variation (see Figure \ref{fig:posing_and_commen}-\circled{A}). 
On the other hand, when reviewing gestures generated by Novobo, teachers personally enacted these gestures, allowing them to observe Novobo through their own actions (see Figure \ref{fig:posing_and_commen}-\circled{4}, \circled{7} and \circled{8}). 
During this process, teachers could notice certain principles underlying the use of gestures. 
For instance, in the scene depicted in Figure \ref{fig:posing_and_commen}-\circled{8}, to examine the ``pointing with my hand'' gesture generated by Novobo, G9P26 performed two variations: one with the index finger and another with a closed palm. Upon observing this, G9P25 commented that using the palm was preferable over the finger, a suggestion also acknowledged by two other teachers.

In sum, teachers initially immersed themselves in the given scenario by envisioning their own teaching situations in the classroom. During the review phase of Novobo's responses, they enact Novobo's gestures, observing Novobo through their own performances.

\subsection{Commentary --- Analysis and Deliberation}

During the commentary phase, teachers evaluated the gestures generated by Novobo. Observation showed that throughout this process, there was extensive discussion among teachers regarding the use of instructional gestures.

\textbf{Vignette.} 
At this stage (see Figure \ref{fig:posing_and_commen}-\circled{C}), the Group 2 teachers collectively addressed the scenario to help students focus their attention, accompanied by the instructional text ``pay attention to here.''
One of Novobo's gestures was: ``Scan the class with your eyes, then gently point to your eyes with a finger, indicating `Look here', to ensure students concentrate.''
After G2P5 tried pointing to her eyes, she felt that this was not an appropriate gesture, and G2P6 agreed, stating, \textit{``We won't point to our own eyes.''} (\textbf{E})
Yet, they both agreed with Novobo's instructional intention that a gesture should be used to capture students' attention. This prompted G2P6 to ask, \textit{``How should we use gestures to attract students' attention?''} (\textbf{S})
She then clapped her hands twice and suggested that clapping twice before speaking could be a way to draw students' attention (\textbf{E}).
Further, she connected this method to clapping as a way of affirming students and remarked, \textit{``This approach could be confused with clapping for encouragement,''} leading her to consider the differences between the two types of claps.
After physically experimenting a few times, she shared: \textit{``This form of clapping, distinct from encouraging applause, is firm and forceful, whereas encouraging clapping is quick and light.''} (\textbf{E})
After observing and imitating this action (\textbf{S}), G2P5 agreed and added that the rhythm of attention-grabbing claps should be slower (\textbf{E}).

\textbf{Knowledge exchange (E\&S).} 
As presented in Figure \ref{fig:esc}, in this phase, teachers mainly express their views on the use of nonverbal behaviors by evaluating the gestures suggested by Novobo (\textbf{E}), while socialization occurs through listening to others' feedback (\textbf{S}). 

Two main scenarios emerge during the commentary process. 
First, teachers physically attempted the gestures generated by Novobo and express disagreement (see Figure \ref{fig:posing_and_commen}-\circled{16}, \circled{17} and \circled{18}). 
This disagreement often prompted them to reflect on how to adjust the gesture for more appropriate use. 
For instance (see Figure \ref{fig:posing_and_commen}-\circled{D}), after G1P1 and G1P3 remarked that tapping the chest to signify ``thinking'' felt inappropriate, G1P1 demonstrated a circular motion with her index finger near her head and said, \textit{``We could do this instead.''} 
This suggestion was endorsed by other teachers.
One interesting phenomenon observed is that when teachers find the gestures generated by Novobo inappropriate, some show a sense of openness and tolerance. 
For example, G1P4 commented, \textit{``Maybe some people do use these gestures; it's just that we’re not used to them.''} 
Similarly, G4P10 remarked, \textit{``It's not that this is wrong, it's just that my perspective is different.''}
This openness may stem from the mentor-mentee dynamics, which fosters a inclusive and supportive atmosphere. 
In the second scenario, teachers agreed with the gestures suggested by Novobo, which often led them to relate these gestures to real classroom situations (see Figure \ref{fig:posing_and_commen}-\circled{20} and \circled{21}). 
For example, when evaluating the gesture of ``forming a telescope shape with hands,'' G5P15 commented, \textit{``For younger students, I think this would be great.''} Such discussions---about tailoring gestures based on age group, subject matter, and individual teaching style---were common during this stage, reflecting teachers' nuanced understanding of how gestures should be used in teaching. These insights were shared and confirmed collectively among teachers through discussion.

Compared to the discussions in the posing question stage, the exchanges in this stage are more in-depth and better reflect what teachers find essential in the teaching process. 
For instance, in Group 6, during the initial stage, G6P18, G6P19, and G6P20 briefly dismissed Novobo's suggestion to ``imitate the shape of ears with hands'' as\textit{ ``far-fetched.''} 
However, during the evaluation stage, their insights were more developed: \textit{``In the classroom, teaching should start with concrete content, followed by imagery, and then transition to abstract concepts. For example, if you're talking about ears, you should first show your own ears...''}

We found that in order to evaluate Novobo, throughout this stage, teachers engaged in in-depth discussions to share and exchange their personal experiences.

\subsection{Demonstration --- Integrating Perspectives}

\begin{figure*}[h]
    \centering
    \includegraphics[width=1\linewidth]{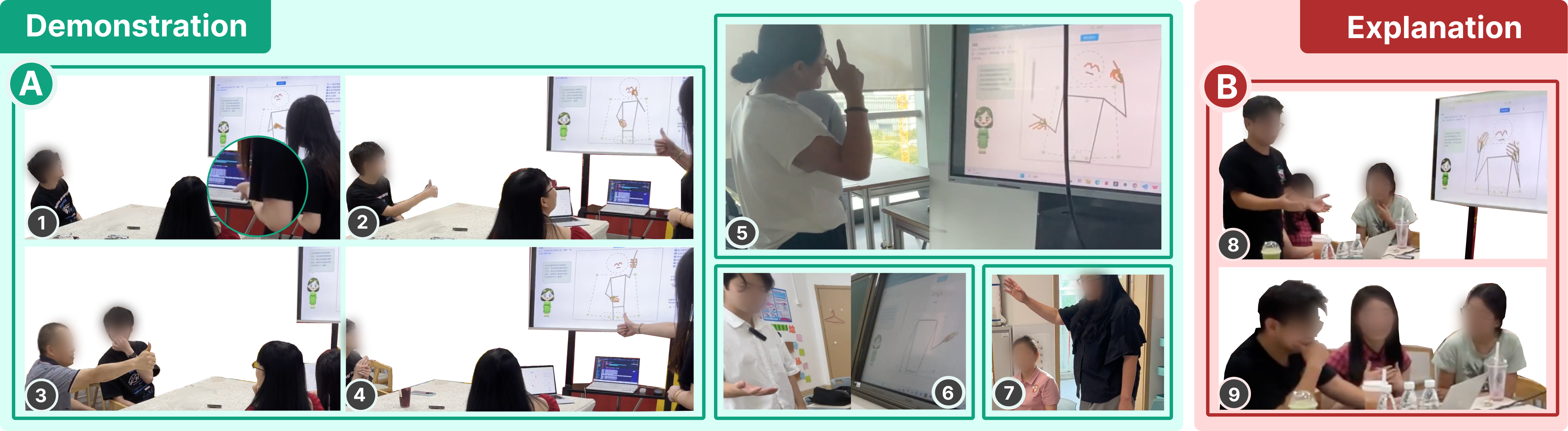}
    \caption{Typical example scenarios of users engaging in the \textit{Demonstration} and \textit{Explanation} stage of the system.}
    \label{fig:demonstration_and_explanation}
\end{figure*}

During the demonstration stage, teachers are required to personally enact instructional gestures they deem suitable for the teaching context and text content. 
We observed that in this stage, teachers not only communicate through verbal means but also engage in communication and offer suggestions through nonverbal behaviors.

\textbf{Vignette.} Figure \ref{fig:demonstration_and_explanation}-\circled{A} provides a typical example of this stage. 
Group 1 collaboratively undertook a teaching scenario that involved praising a student, using the phrase, ``Your vocabulary is really rich.'' 
To better demonstrate for Novobo, G1P1 first practiced her gestures in front of the skeletal mirror (\textbf{E}). 
While saying \textit{``Your vocabulary is really rich,''} she placed both hands near her abdomen and clapped (see Figure \ref{fig:demonstration_and_explanation}-\circled{1}).
Observing this (\textbf{S}), G1P3 and G1P4 suggested she raise her hands higher. G1P4 stated, \textit{``You could raise your hands to the side of your cheek,''} and demonstrated a clapping gesture with both hands positioned near the left side of the cheek (\textbf{E}).
When G1P1 tried clapping with raised hands, she hesitated. G1P4 added that if this gesture felt uncomfortable, she could instead \textit{``give a thumbs-up.''} 
G1P1 accepted this suggestion and practiced again (see Figure \ref{fig:demonstration_and_explanation}-\circled{2}).
At this point, G1P2 added, \textit{``Don't give yourself a thumbs-up; do it towards the students,''} while demonstrating the gesture by extending the thumb outward from in front of the chest (\textbf{E}) (Figure \ref{fig:demonstration_and_explanation}-\circled{3}).
In reponse, G1P1 practiced again and finalized her gesture (\textbf{C}) (see Figure \ref{fig:demonstration_and_explanation}-\circled{4}).

\textbf{Knowledge exchange and co-construction (E\&S\&C).} 
As shown in Figure \ref{fig:esc}, in this stage, teachers mainly externalized (\textbf{E}) their knowledge by demonstrating gestures to Novobo or offering suggestions to other teachers (see Figure \ref{fig:demonstration_and_explanation}-\circled{5}, \circled{6} and \circled{7}). 
At the same time, they underwent a process of socialization (\textbf{S}) by observing their colleagues' actions and listening to their suggestions. 
This process of externalization and socialization often occured through a combination of verbal communication and physical gestures. 
For instance, G6P18 observed that G6P19 raised her hand high and then lowered it when emphasizing a particular concept. G6P19 acknowledged this, remarking that she had never noticed this habit before (see Figure \ref{fig:demonstration_and_explanation}-\circled{7}). 
This process could make the underlying principles of instructional gesture use more explicit for the teachers.
Before demonstrating, teachers continuously experimented with their gestures and reflect on them by reviewing skeletal feedback of their movements. 
For example, after watching a replay of his gestures, G10P30 felt that the range of his movement could be larger, leading him to re-record his gestures (see Figure \ref{fig:demonstration_and_explanation}-\circled{6}). 
This self-reflection process, aided by the skeletal replay, allows teachers to adjust their gestures as they see themselves in action.

Differing from the previous two phases, this stage required teachers to finalize and deliver a specific instructional gesture to Novobo. Therefore, teachers needed to reach a consensus, collaboratively constructing their shared understanding of nonverbal knowledge (\textbf{C}). 
As described in the vignette above (see Figure \ref{fig:demonstration_and_explanation}-\circled{A}), G1P1 resonated with the idea of ``expressing towards the students'' and ultimately developed her own teaching gesture based on that feedback.

In summary, demonstration phase advanced the practice of teachers expressing their understanding both verbally and bodily, promoting the exchange and co-construction of embodied knowledge through multi-modal means. 

\subsection{Explanation --- Collective Meaning-making}

The final stage involved the explanation of instructional gestures demonstrated in the previous phase. In this stage, teachers jointly deliberated on the principles and reasons behind their use of gestures, articulating these insights for collective meaning-making.

\textbf{Vignette.} 
The teachers in Group 4 collaboratively addressed a teaching scenario centered on experiencing nature, paired with the teaching text, ``In nature, we discover beauty and feel beauty.'' (Figure \ref{fig:demonstration_and_explanation}-\circled{B})
During the previous demonstration stage, G4P10, while saying ``feel beauty,'' extended his hands outward, palms facing up.
In the explanation phase, G4P11 questioned, \textit{``Why do the hands move outward when feeling beauty?''} She then suggested that feelings should be expressed inward. At this point, she demonstrated by crossing her hands over her chest. Upon reflection, G4P10 explained that his gesture was intended to \textit{``Encourage students to express their inner feelings outward.''}
G4P11 nodded in agreement, acknowledging the explanation, and G4P10 subsequently entered this consensus principle into the system.

\textbf{Knowledge exchange and co-construction (E\&S\&C).} 
As illustrated in Figure \ref{fig:esc}, in this phase, the teachers primarily utilized verbal communication to externalize (\textbf{E}) and socialize (\textbf{S}) knowledge, ultimately integrating it by inputting the formed consensus into the Novobo system (\textbf{C}). 
We observed that when Novobo invited teachers to explain their gestures, many initially responded with phrases like \textit{``I just did it casually''} (G4P10) or \textit{``there wasn't any particular reason''} (G5P15). 
However, once the teachers engaged in more in-depth discussion, they revealed integrated different underlying considerations. 
For example, in Group 2, when G2P6 explained her gesture of extending the thumb to encourage students, she mentioned that the goal was to make the gesture \textit{``simple and easy to understand.''} 
Building on this, G2P5 added that this gesture could also\textit{ ``stimulate students' enthusiasm for learning through affirmation.''} 
Ultimately, G2P6 integrated both viewpoints and entered the following into the system (\textbf{C}): using the thumb to affirm students is a widely recognized and easily understood gesture that can stimulate students' motivation to learn.
Moreover, it is worth noting that the teachers expressed great interest and approval when reviewing Novobo’s summary of the instructional gesture principles derived from their inputs.

In this stage, teachers primarily share and discuss the principles and considerations behind the use of nonverbal behaviors. 
Through this process, implicit embodied knowledge is ultimately made explicit, discussed, and co-constructed, leading to a shared understanding within the group.

\section{Results for RQ2: Teachers' Experiences about Novobo's Core Design Features}

\subsection{DF1: Well-Referenced Domain Knowledge --- \textit{``I tend to trust theoretical knowledge provided by Novobo more.''}}

Teachers frequently took the initiative to consult the knowledge resources provided during system use. Access to theoretical knowledge enhanced their learning and heightened their awareness of the importance of instructional gestures.
As G1P4 noted, \textit{``Understanding the sources of knowledge and learning these concepts has been quite enlightening for us.''} G10P29 mentioned that these knowledge sources made him realize that nonverbal behavior was an area for improvement: \textit{``I was very surprised. I didn't realize there was so much research in this field; it might have been a blind spot for me.''}

Teachers acknowledged that they had previously overlooked the use of gestures and recognized the need to place greater emphasis on this aspect in the future. As G2P6 noted: \textit{``Typically, I haven't focused much on using gestures. It was only when we came here for the testing that I started organizing these elements.''} G5P13 emphasized the need to pay special attention to nonverbal teaching behaviors in future teaching. 
G4P11 admitted to a lack of reflection in the past: \textit{``Sometimes when I use gestures, it's out of habit rather than conscious awareness.''} 
G7P20 discussed the importance of understanding the meaning behind gestures: \textit{``If I knew its meaning, I would place more emphasis on that gesture. I would use it more to encourage students, boost their confidence, or make myself more approachable.''} 
Moreover, the presentation of knowledge sources increased teachers' trust in the AI-generated results. As G10P28 stated, \textit{``Colleagues' advice is often based on personal experience, whereas Novobo's recommendations are more grounded in theory and literature. I tend to trust theoretical knowledge provided by Novobo more.''}

These findings suggest that in the process of  teachers' professional development, providing both AI-generated suggestions and the theoretical basis for these suggestions effectively increases the explainability of AI.  This not only enhances their learning of relevant knowledge but also increases their trust toward AI-generated responses.

\subsection{DF2: Embodied Practice with a Skeletal Mirror --- \textit{``Viewing the skeleton feels less stressful.''}}

Teachers' gestures represent a form of embodied, tacit knowledge that cannot be easily expressed through language or text. Therefore, in order to facilitate the sharing of this implicit knowledge, it is essential to provide teachers with opportunities to express themselves physically.
During their use of the system, embodied practice occurred frequently. 
In the Demonstration stage, teachers practiced and demonstrated gestures in front of the skeletal mirror, which encouraged self-reflection through a process of observing and doing simultaneously. This allowed them to externalize and socialize this form of embodied, tacit knowledge through continuous practice across these stages.

Feedback based on skeletal movements enabled teachers to reflect on their practices by mirroring their actions during exercises, without fostering self-criticism or anxiety about being judged by others. For example, in Group 5, when G5P14 realized she needed to turn on the camera, she initially displayed noticeable nervousness and hesitation. However, once she discovered that the system would only capture her skeletal outline rather than her full appearance, she became more relaxed and willing to perform various actions. She noted, \textit{``Before I knew it was only recording the skeleton, I thought it would capture my appearance, and I felt pressured. But realizing it's just a skeleton made it much better.''} Participants also highlighted that this design allowed them to focus more on their non-verbal behavior rather than their appearance. G9P25 remarked, \textit{``It completely focuses my attention on my body movements, so I won't be looking at my face or anything.''}

By demonstrating their own gestures in front of peers, teachers moved from individual practice to collective reflection. Teachers externalized their individual nonverbal knowledge through bodily expressions during interacting with the skeletal mirror. After reflecting and comparing, other observing teachers would offer suggestions and share their understanding of teaching gestures, either through actions or verbal comments, as shown in Figure \ref{fig:demonstration_and_explanation}.

This process of reflecting on gestures and embodying tacit knowledge through physical expression not only aids self-awareness but also facilitates the transfer and co-construction of nonverbal instructional techniques.

\subsection{DF3: Mentor-mentee Dynamics between Teachers and AI --- \textit{``Evaluating AI comes with no psychological burden.''}}

Teachers distinguished between guiding AI and having AI serve as an ``expert teacher'' to guide them, with most expressing skepticism about the latter. 
As G3P8 stated, \textit{``If it acts as an `expert teacher', I might not fully trust its data because I have experienced issues with large language models before.''} However, as an apprentice, AI is more likely to facilitate personal reflection.
As G5P14 put, \textit{``I think teaching AI is more of a self-reflective journey, because in reality, you wouldn't ponder many issues unless you specifically focus on them.''}

Teachers noted that the psychological burden of interacting with a teachable agent is lower. 
Firstly, teachers shared that pointing out issues in other teachers' teaching methods might involve social considerations, whereas evaluating AI does not carry such concerns. For instance, G10P29 mentioned, \textit{``Privately, I might be blunt, but in formal evaluations, I tend to offer only positive feedback.''} This is consistent with previous research \citep{seroussi2019reflections}.
Moreover, less experienced teachers expressed a lack of confidence in evaluating others, with one teacher saying, \textit{``Usually, it's others who give me feedback; I seldom give feedback to others. My experience is too limited.''} 
In contrast, when interacting with the AI-based agent, G5P13 mentioned, \textit{``Because when evaluating AI, I don't attribute any characteristics to it, which lets me focus purely on the issue itself, like instructional gestures.''} This finding aligns with previous research, indicating that teachers can express themselves more freely and offer deeper insights when assessing virtual entities \citep{brown2004teachers}. 
Teachers also expressed the idea of growing together with AI, as G1P3 mentioned, \textit{``Perhaps you (AI) can draw from my experience, and I can see what new methods you (AI) have.''} G9P26 also expressed, \textit{``This back and forth interaction feels good, like having grown from just one attempt, and then progressively, allowing it to grow while I also learn a lot.''} 

In summary, facilitating teachers' peer learning through a teachable agent could reduce peer pressure, promote an inclusive atmosphere during the interaction with AI, and potentially promote a growth mindset (co-learning with the AI agent). 

\section{Discussion}

\subsection{Summary of Research Questions}

In addressing \textbf{RQ1}, we found that throughout the four stages of using the system, teachers continuously externalized their tacit embodied knowledge, engaged in socialization, and combined their insights (see Figure \ref{fig:esc}).
In the \textit{Posing Question} stage, teachers engaged with potential instructional gestures by performing and mimicking both Novobo's and their peers' gestures (\textbf{E\&S}).
During the \textit{Commentary} stage, teachers evaluated Novobo's  gestures through personal assessments and shared insights during group discussions (\textbf{E\&S}).
In the \textit{Demonstration} stage, teachers practiced and recorded their instructional gestures, incorporating peer feedback  (\textbf{E\&S}) and achieving consensus through collaboratively embodied practice (\textbf{C}).
In the \textit{Explanation} stage, verbal discussions helped teachers externalize and share their understanding  (\textbf{E\&S}), integrating their collective insights into the Novobo (\textbf{C}). 

These results provide new insights into the key role a teachable AI agent can play in facilitating the externalization of tacit embodied knowledge, as well as the socialization and combination of teachers' insights during professional development group discussions. Using TAs could complement existing strategies aimed at helping teachers develop skills for effective instructional gesturing like peer assessments \citep{brown2004teachers}, video analysis \citep{hamidah2019video} and behavioural change interventions \citep{mcdonald1973behavior}. This also highlights the potential for extending Learning by Teaching (LBT) and Teachable Agent-based approaches beyond student support, as explored in most previous research \citep{song2017designing,johansson2023intelligent}, to support teacher development. Specifically, it shows promise for enhancing peer learning and the development of instructional gesture strategies by integrating the TA into existing practices \citep{khuman2024impact}.

In addressing \textbf{RQ2}, we discovered that teachers appreciated Novobo's answers with well-referenced domain knowledge (\textbf{DF1}) not only for their educational value in knowing more about the importance of instructional gestures based on theoretical knowedlge \citep{ goldin2011learning, roth2001gestures, smotrova2017making, novack2015learning}, but also for increasing trust in the responses produced by LLMs. 
For the aspect of embodied practice with a skeletal mirror (\textbf{DF2}), teachers reported less stress and a heightened willingness to use physical expressions. 
Regarding the support of learning through mentor-mentee dynamics between teachers and AI (\textbf{DF3}), teachers felt that this narrative approach mitigated social judgment, a typical concern within professional development practices \citep{seroussi2019reflections,xie2008effect}, allowing them to express their opinions more deeply. This finding is consistent with prior research, suggesting that teachers are more open and insightful when evaluating virtual entities \citep{brown2004teachers}. Moreover, we found that teachers indeed appreciated the potential of the LBT approach as a way to provoke reflection on their own teaching practices and collectively constructing meaning, which coincides with the aim of this pedagogical approach \citep{jin2023teach,choi2021reconsidering}.

In the rest of this section, we discuss more in depth the design implications from these findings that could benefit future HCI and design research and practice.

\subsection{Supporting Teachers' Free Expression Through AI-based Mentor-Mentee Narrative}

Existing research has indicated that study groups, as a critical process in teacher professionalization, face challenges such as the social pressure teachers experience when offering criticism to colleagues, which can hinder the depth and effectiveness of feedback\citep{brown2004teachers}. 
Similarly, research indicates that social comparison should largely be avoided unless it is combined with other, more informative feedback \citep{weidlich2024emotional}. 

HCI research has explored the use of LLMs as learning peers to enhance student engagement in the classroom. However, when the AI classmate displays more expertise than the user, it may lead to user anxiety \citep{liu2024classmeta}. Another study indicates that positive emotional expressions from AI team members can affect human teammates’ trust and collaboration \citep{mallick2024you}.
Similarly, \citet{nguyen2023role} emphasize that conversational agents with intentional role designs, such as less-knowledgeable peers or expert mentors, can foster productive discussion and systems thinking in group settings. 
These findings suggest that varying role designs impact participation dynamics and learning outcomes. 
Our evaluation also revealed that when the AI agent assumes the role of a mentee receiving feedback and evaluation, teachers are able to express their insights more comprehensively and profoundly, with fewer anxiety. This finding suggests that within teacher study groups, a teachable agent not only supports learning through the ``learning by teaching'' approach, as discussed in other HCI studies \citep{jin2024teach}, but also serves as a buffer for exchanging critical feedback.
In addition to lowering the social barriers to expressing critical viewpoints, teachers also display an inclusive attitude towards the diverse perspectives generated by the teachable agent. 
Therefore, we believe that teachable agents hold significant potential in fostering an inclusive and constructive atmosphere within teacher professional learning groups, warranting further exploration and design.

\subsection{Supporting Community Knowledge Co-Construction and Sharing Through Collaborative Teaching}

In professionalization, much of the knowledge-in-practice lacks a universal standard \citep{cochran1999chapter}. 
This type of highly personalized and experiential knowledge, which is heavily dependent on context and local culture, is difficult to acquire solely through theoretical learning. 
Instead, it requires practitioners to immerse themselves in a shared practice environment, where they gradually acquire it through socialization \citep{wenger1999communities}. 
Each local practitioner community develops a unique appreciative system for practical knowledge. 
However, teachers in the initial design study with teachers using the technological probe (presented in Section \ref{formative}), expressed that during their early career stages, they could only learn fragmented insights about instructional gestures from their mentors. 
Therefore, communities should create a space for practitioners to exchange, share, and learn from each other, supporting group reflection and the co-construction of updated shared knowledge and evaluation systems.

Collaborative problem-solving in peer interaction is key to transmitting tacit knowledge within the community\citep{ojeda2024qualitative}. 
Supporting this, \citet{de2024assessing} argue that LLM-based chatbots enhance group awareness—a critical element for effective collaboration—by making individual goals and actions transparent to the group. 
\citet{borge2018learning} have demonstrated that requiring individuals to provide evidence for reflective assessments can deepen the knowledge gained from collaborative processes. 
\citet{hung2011influence} pointed out that “reputation feedback” positively impacts the quality and quantity of knowledge-sharing, while altruism enhances the satisfaction derived from sharing. By designing a mentee AI-agent in need of guidance, we aim to stimulate teachers’ altruism, while Novobo provides reputation feedback by affirming and summarizing the knowledge shared by teachers.
In line with this, we observed that when users provided feedback based on Novobo's performance, they tended to produce more insightful reflections. 
When teachers evaluated Novobo's gestures, they externalized their tacit knowledge—knowledge that they had not yet deliberately considered—within the group. 
Simultaneously, Novobo, as a medium that encapsulates group knowledge, allows teachers to become aware of and reflect on each other’s nonverbal teaching behaviors. 
When teachers engaged in thorough discussions and jointly decided to base Novobo on what they collectively considered the best example, and then provided a shared interpretation, the co-construction of knowledge and the co-shaping of values occurred within the local practitioner community. 
We suggest that other researchers and designers could document relevant teaching considerations reflected in the interaction process with digital systems, forming a teaching profile of an individual teacher, a group, or even an entire teaching organization. 
This would not only promote real-time knowledge exchange and sharing but also foster the co-construction of knowledge within the community. 
Additionally, given the potential system’s 24/7 availability, it is worth exploring the possibility of teachers using the system asynchronously to facilitate knowledge exchange.

\subsection{Enhancing the Transparency of Generated Information Using AI in Education Contexts}

For highly specialized learning situations, such as generating appropriate nonverbal behaviors for teachers, general LLMs and common-sense reasoning are insufficient and are often not trusted by professionals \citep{harasta2024cannot}. 
Many studies have emphasized the importance of using explainable AI \citep{shen2023towards, kaadoud2022explaining}, addressing fields such as education \citep{khosravi2022explainable, hostetter2023xai, yadollahi2024extra, conati2018ai,xu2023transparency}, healthcare \citep{rubegni2024designing,kim2024stakeholder,seitz2022can} and other contexts \citep{dikmen2022effects,kim2023designing}. 

Employing the RAG \citep{gao2023retrieval} approach to provide domain-specific knowledge to LLMs has proven beneficial in addressing these highly specialized learning situations \citep{li2024enhancing,yan2024vizchat}. 
In our study, we not only presented the results generated by the large model but also provided the relevant knowledge and intermediate reasoning steps retrieved during the RAG phase—such as instructional intentions. 
Our findings show that this approach not only helps teachers learn the relevant theoretical knowledge but also reduces their skepticism toward AI-generated results.
Building on these insights, we advocate for a more nuanced application of AI tools in educational settings. Moreover, \citep{gomez2023mitigating} found that user-AI interaction in knowledge creation significantly increases trust in AI knowledge. In our study, users first review AI-generated gesture suggestions and domain knowledge, then refine these gestures based on their practical experience, fostering human-AI collaboration, reducing distrust, and deepening reflection on teaching gestures. Other researchers and designers can go beyond simply delivering final outputs by providing detailed explanations, reasoning processes, and supporting evidence. 
Such transparency is essential to cultivate trust and efficacy in AI applications, ensuring they meet the intricate demands of educational professionals. At the same time, it provides users with the opportunity to some extent make independent judgments based on the provided materials.

\subsection{Facilitating Embodied Practice in Nonverbal Competence Training}

Previous research has demonstrated that nonverbal behaviors constitute a form of tacit knowledge, acquired through practice and challenging to communicate via textual or verbal methods \citep{westerlund2021s}. Therefore, we incorporated opportunities for embodied practice to facilitate the development of instructional gestures.
Our findings indicate that in nonverbal competence learning, teachers' embodied interactions can serve as a means to externalize their embodied skills, facilitating the exchange and co-construction of embodied knowledge. 
For example, when teachers physically experiment with instructional gestures they might use in certain situations, their internal tacit knowledge becomes externalized through embodied actions and is received by others.
Moreover, we found that displaying the body through a skeletal representation, rather than showing the full body, effectively reduced teachers' self-consciousness. 
Teachers were more willing to experiment with different instructional gestures in front of the skeletal mirror and reflect on their recorded performances, which encouraged them to further practice and refine their gestures.

We recommend that researcher and designer interested in this are to explore further the provision of opportunities for collaborative embodied practice and reflection in nonverbal skills training, such as instructional gestures, nonverbal communication skills in other areas such as  nursing, and mechanical operation skill development for workers. 
Moreover, supporting the development of embodied skills and tacit knowledge in groups can be enhanced by grounding the process in established theoretical frameworks, such as the SECI model \citep{nonaka1994dynamic, westerlund2021s, mendoza2022assessing}. An area that has yet to be explored is not only how people externalise tacit knowledge but also how the internalisation of such embodied tacit knowledge occurs and how it can be effectively supported.

\subsection{Limitations and Future Work}

Our work has several limitations. 
Firstly, due to the current constraints of generative AI, Novobo conveys the generated gestures to teachers through text. This method, being relatively indirect and requiring interpretation from the teachers, may lead to variances in how different educators perceive and decode these gestures.
Future work can explore more direct and intuitive methods for conveying gestures, such as integrating multimodal outputs like visual or animated representations. This could help reduce interpretation variability and ensure more consistent understanding of the gestures across educators. Additionally, incorporating real-time feedback or embodied AI agents could further enhance the interaction, making the system more aligned with the embodied nature of instructional gestures. 

Secondly, our research participants consist exclusively of teachers from first-tier cities, who exhibit a greater familiarity with employing AI in educational contexts. Teachers operating in varied settings might engage differently and hold distinct perspectives towards an LLM-driven teachable agent. Future work can explore the experiences and perceptions of educators from diverse geographic and socio-economic backgrounds to better understand how factors such as access to technology, educational resources, and AI literacy influence their interaction with AI systems. This could lead to the development of more inclusive, adaptable teachable agents that accommodate the varying needs and constraints of educators in different environments.

Finally, the current approach necessitates synchronous communication among teachers. However, given their busy schedules, it might be challenging for them to find common times to engage in such activities. Future work could explore how a teachable agent might serve as a medium for the exchange and construction of embodied knowledge among teachers in an asynchronous environment.

\section{Conclusion}


In this study, we designed Novobo, a teachable AI agent aimed at supporting teachers in collaboratively learning instructional gestures.
We utilized prompt engineering and Retrieval-Augmented Generation (RAG) techniques to embed both theoretical and practical knowledge of instructional gestures into large language models (LLMs), enabling the generation of appropriate gestures while also providing well-referenced knowledge.
Novobo encourages teachers to evaluate its generated gestures and to provide demonstrations through both verbal and bodily inputs. The use of a skeletal mirror to display embodied practices was considered to reduce anxiety stemming from self-consciousness or being observed by others.
Our findings also showed that positioning the AI as a mentee was seen as effective in reducing social pressures, fostering collaborative learning, and promoting in-depth reflection.
Moreover, the collaborative process of teaching Novobo supported teachers externalize, exchange, and internalize their tacit knowledge, promoting the co-construction of instructional gesture knowledge within the local teaching community.
Overall, our research highlights the potential of teachable AI agents to serve as a medium for knowledge dissemination, offering valuable insights for the design of future teachable AI systems.

\section{Declaration of generative AI and AI-assisted technologies in the writing process}

During the preparation of this work the authors used ChatGPT(GPT-3.5) in order to improve language in a few parts in this paper. After using this tool/service, the authors reviewed and edited the content as needed and take full responsibility for the content of the publication.


\bibliographystyle{elsarticle-harv} 
\bibliography{cas-refs}





\end{document}